\documentclass[12pt]{article}
\usepackage[papersize={200mm,310mm},tmargin=15mm,bmargin=15mm,lmargin=15mm,rmargin=15mm]{geometry} 

\usepackage{graphicx}

\usepackage{longtable}

\begin{document}

\title{ Fractional cable equation for  general geometry, a model of axons with swellings and anomalous diffusion}
\author{ Erick J. L\'opez-S\'anchez $^{(1)}$\thanks{lsej@unam.mx}, 
 Juan M. Romero $^{(2)}$ \thanks{jromero@correo.cua.uam.mx},  Huitzilin Y\'epez-Mart\'inez$^{(3)}$ \thanks{huitzilin.yepez.martinez@uacm.edu.mx }\\[0.5cm]
\it $^{(1)}$Posgrado en Ciencias Naturales e Ingenier\'ia, \\
\it Universidad Aut\'onoma Metropolitana, Cuajimalpa. \\
\it Vasco de Quiroga 4871, Santa Fe Cuajimalpa, D. F. 05300 M\'exico.\\
 \it $^{(2)}$Departamento de Matem\'aticas Aplicadas y Sistemas,\\
\it Universidad Aut\'onoma Metropolitana-Cuajimalpa,\\
\it M\'exico, D.F  05300, M\'exico\\
\it $^{(3)}$Universidad Aut\'onoma de la Ciudad de M\'exico, \\
 \it Prolongaci\'on San Isidro 151,  San Lorenzo Tezonco, Iztapalapa,  \\
 \it Ciudad de M\'exico 09790, M\'exico. \\}

\date{}

\pagestyle{plain}

\maketitle

\begin{abstract}

Different experimental studies  have  reported anomalous diffusion in brain tissues and notably this anomalous diffusion is expressed through fractional derivatives. Axons are important  to understand   neurodegenerative diseases such as multiple sclerosis, Alzheimer's disease and  Parkinson's disease.  Indeed,  abnormal accumulation  of proteins and organelles in axons  is a hallmark feature of these diseases. 
The diffusion in the axons can become to anomalous as a result from this abnormality. In this case  the voltage propagation in axons is affected. 
Another hallmark feature of different neurodegenerative   diseases is given by discrete  swellings along the axon. In order to model the voltage propagation in axons with anomalous diffusion and swellings,  in this paper we propose a fractional cable equation for  general geometry. This generalized  equation depends on fractional parameters and geometric quantities such as the  curvature and torsion of the cable. 
For  a cable with a constant radius we show that  the voltage  decreases when the fractional effect  increases. 
In  cables with swellings  we  find that  when the fractional effect 
 or  the  swelling radius increase, the voltage decreases. A similar behavior is obtained when the number of swellings  and the fractional effect increase.  Moreover,  we find that when  the radius swelling (or the number of swellings) and  the fractional effect increase  at  the same time, the voltage dramatically decreases.  

\end{abstract}
\section{Introduction}

In  biological organisms there are hydrogen atoms  in abundance, particularly in water and fat.  
These atoms  allow to study  biological organisms with non invasive techniques, such as Magnetic Resonance Imaging (MRI). 
Indeed,  using a magnetic field and the  Zeeman effect,  the diffusion process of water molecules   in biological tissues can be mapped \cite{NMR1,NMR2}.
Thus, using these techniques it is possible to study some physiological properties of biological tissues, for example water molecules diffusion patterns can  reveal microscopic details about tissue architecture and can reflect interactions with many obstacles, such as macromolecules, fibers, and membranes \cite{NMR1,NMR2,NMR3,NMR4}. Notably, in neuroscience  water diffusion can provide  information
about white matter integrity, fiber density, uniformity of nerve fiber direction,   axonal membranes and cytoskeleton properties, etc \cite{NMR3,NMR4}. 
In fact,  water diffusion has been used  to detect and characterize different neurodegenerative  diseases \cite{rvd}. For instance, using Diffusion Tensor Imaging (DTI), altered diffusion has been detected  in white matter of  subjects  with multiple sclerosis \cite{sclero1,sclero2,sclero3}.  Also, using DTI, white matter alterations were found in the corpus callosum of subjects with Huntinton's disease \cite{rosas}. Moreover, using DTI,
relevant  white matter abnormalities were found in Alzheimer's disease \cite{ad1,ad2}.  In addition,  using MRI, hippocampal atrophy has been detected in Parkinson's disease \cite{pk1}. Even more, different experimental studies have shown  that water  diffusion in some tissues can not be described by a Gaussian model, but 
 as an anomalous diffusion expressed through fractional  calculus \cite{frac-1,frac-2,frac-3,frac-4,Be03,Be06},
in particular in  brain tumours \cite{brain-t1,brain-t2,brain-t3} and dendrites \cite{f3}.  Furthermore,  some authors have suggested that the anomalous diffusion  parameters might  play a role in the diagnosis  of brain diseases \cite{Qin}. It is worth mentioning that in different neurodegenerative diseases are reported  abnormal accumulations of proteins and organelles which generate  disruption axonal transport \cite{transport1,transport2,transport3}. Additionally, different theoretical studies support  the claim  that anomalous diffusion appears in  a  heterogeneous medium \cite{f1,hm1,hm2}. For these reasons,   diffusion and anomalous diffusion in brain tissues are relevant   
in the study  of  the brain physiology. Other studies about anomalous diffusion in cellular system can be seen in \cite{adc1,adc2,adc3,adc4}.\\

Another essential aspect to understand how the brain works is given by its electrical activity. 
Thus, in order to obtain a reasonable  model of the brain, it is important to know how the voltage propagates 
in brain tissues  with anomalous diffusion.  In particular, due  that  axons are crucial in neuron-to-neuron communications and those 
can be described  as cables, we should  know how the voltage propagates in a cable with anomalous diffusion. In this respect,   to study the voltage $V(x,t)$ in a
straight cylindrical  cable   with a circular cross-section of constant diameter $d_{0}$ and  anomalous diffusion, 
recently  some authors have  proposed a fractional cable equation as follows \cite{f5,f6}
\begin{eqnarray}
c_{M}\frac{\partial V(x,t) }{\partial t}=  \beta_{\nu} D_{\nu t} \left(  \frac{d_{0} }{4r_{ L} }  \frac{\partial ^{2}V(x,t) }{\partial x^{2}}-   i_{ion}\right),  
\label{frac1}
 \end{eqnarray}
where $c_{M}$ denotes the specific membrane capacitance, $r_{L}$ denotes the
longitudinal resistance and $i_{ion}$ is the ionic current  per unit  area into and out of the cable,
 \begin{eqnarray}
D_{\nu t} = \frac{\partial^{1-\nu} }{\partial t^{1-\nu} }, \quad \nu={\rm constant }, \qquad 0\leq \nu \leq 1,
 \end{eqnarray}
is the Riemann-Liouville fractional operator \cite{loke}  and $\beta_{\nu}$ is a constant with $(time)^{1-\nu}$ dimensions. 
The passive cable case, namely when    $i_{ion}=V/r_{M}$ (where  $r_{M}$ is 
 the  specific  membrane resistance)  was  used to study  electrodiffusion of ions in nerve cells \cite{f5,f6}.\\

Additionally,  there are different physiological phenomena where the geometry is important. For instance, 
 axons with non trivial geometry are important  to understand some neurodegenerative diseases, indeed discrete swellings along the axons  appear in  neurodegenerative diseases such as  Alzheimer's disease, Parkinson's disease, HIV-associated dementia and multiple sclerosis.  In fact, 
axons with a  diameter  of approximately  $1 \mu m$ with  a swelling with diameter of approximately $5 \mu m,$  are reported in Parkinson's disease 
 \cite{Ga99}. In addition,  axons with a diameter of approximately   $4 \mu m$
with a  swelling with a diameter of approximately $60 \mu m$ are reported in   multiple  sclerosis  \cite{Tr98};  axons with a diameter  of approximately  $1.5 \mu m$ with  a swelling train, where the swelling diameter varies between $4 \mu m$ and $10 \mu m,$ are reported in Alzheimer's disease  \cite{Xi13, Jo13, Kr13};  axons with a diameter of approximately $6\mu m$ and swellings   with a diameter of approximately  $43 \mu m$  are reported 
in HIV-associated dementia \cite{Bu87, Bu871,Bu872,Bu873,Bu874}.  Other  sizes of the axonal  swellings can be see in \cite{maia1,Sh98}.
Some theoretical studies on cables with non cylindrical geometry  can be seen in \cite{tre3,vetter,bird,kandel,fiala,poznanski1,maia,erick}.
Because   the equation (\ref{frac1}) only describes axons  with cylindrical  geometry, therefore, in order to study the voltage propagation in a cable with  non trivial geometry and anomalous diffusion, this  equation should be generalized.\\

In this paper, we study the voltage propagation in a cable with anomalous diffusion and non trivial geometry. For this purpose, we introduce a fractional cable equation for  a general geometry.  This generalized  equation depends on fractional parameters $\beta_{\nu}$ and  $\nu$ and geometric quantities such as the  curvature and torsion of the cable. For different cable geometries, we show  that  with regard to model a system where the voltage decreases, we should  suppose that $\beta_{\nu}$ increases when $\nu$ decreases. For  a straight cylinder  with a constant radius we show that  the voltage depends on neither the
curvature nor the torsion of the cable and it decreases when $\beta_{\nu}$ increases and  $\nu$ decreases. In addition, cables with swellings are studied. In these last cable geometries we  find that  when the swelling radius increases or  $\beta_{\nu}$ increases and  $\nu$ decreases, the voltage decreases dramatically.  \\

 This paper is organized as follows: in the section \ref{se:fcg} we propose a  fractional cable equation with a general geometry; in the section \ref{se:gc} we analyse  some general properties of the generalized  fractional cable equation; in the section \ref{se:ccf} we consider  the cylindrical cable  with  a constant radius; in the section \ref{se:ns} we study  cable  with swellings.  Finally, in the section \ref{se:s} a summary is given.\\

\section{Fractional cable equation in a general geometry }
\label{se:fcg}

It is well known that to study  the  geometric 
properties of a three dimensional curve $\vec \gamma$ the 
arc length  parameter 
\begin{eqnarray} 
s=\int_{0}^{x}\sqrt{\frac{d\vec \gamma(\zeta) }{d\zeta }\cdot \frac{d\vec \gamma(\zeta) }{d\zeta }} d\zeta, \label{affine}
\end{eqnarray}
is a friendly  parameter. Indeed, 
 using the arc length parameter (\ref{affine}) we can construct the vectors of the Frenet-Serret  frame \cite{docarmo}
\begin{eqnarray} 
\frac{d \vec \gamma (s)}{ds}=\hat T, \qquad \hat N= \frac{\frac{d \hat T}{ds}  }{ \left| \frac{d \hat T}{ds}\right | }, \qquad 
\hat B=\hat T\times \hat N, 
\end{eqnarray}
where $\hat T$ is the unit vector tangent,  $\hat N$ is the normal unit vector and $\hat B$ is the binormal unit vector to the curve. Furthermore, 
using the arc length and the Frenet-Serret  frame,   the Frenet-Serret formulas  can be obtained as follow \cite{docarmo}
\begin{eqnarray} 
\frac{d\hat T}{ds}=\kappa \hat N,\quad 
\frac{d\hat N}{ds}=-\kappa \hat T+\tau \hat B,\quad 
\frac{d\hat B}{ds}=-\tau \hat N,
\end{eqnarray}
 where  $\kappa, \tau$ are the curvature and torsion of the curve $\vec \gamma,$ respectively. \\

\begin{figure}[b]
  \begin{center} 
 \includegraphics[width=.7\textwidth]{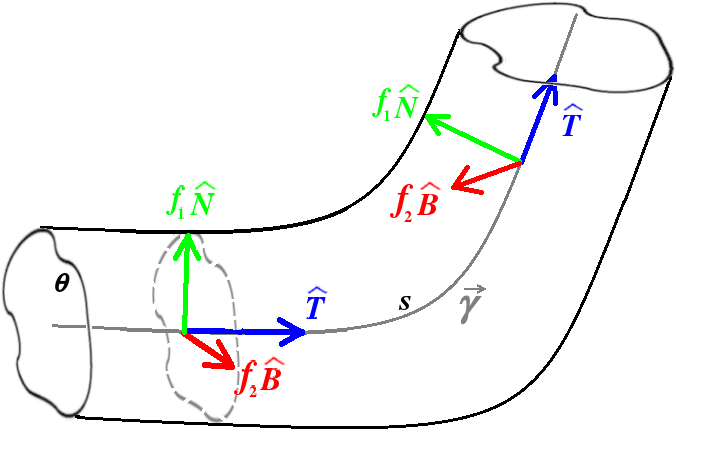}
 \end{center}
  \caption{ Cable with general geometry. The vectors  $\hat T, \hat N, \hat B$ are shown in two different points on the curve $\vec \gamma.$}\label{tubo}
     \end{figure}

We can employ the Frenet-Serret  frame to construct a cable model. Actually, 
we can propose  a general cable as the region bounded by the following surface 
\begin{eqnarray}
\vec \Sigma(\theta,s)=\vec \gamma(s)+f_{1}(\theta,s)\hat N(s)+f_{2}(\theta,s) \hat B(s), \label{surface}
\end{eqnarray}
where $\theta$ is an angular variable. Notice that employing the angular coordinate $\theta,$ the functions $f_{1}(\theta,s), f_{2}(\theta,s)$ and the vectors  
$\hat N(s), \hat B(s)$ we are constructing the cable over the curve $\vec \gamma(s).$  In  Fig. \ref{tubo} we can see a representation of the surface (\ref{surface}).  For instance,  a cable with  a deformed circular cross-section, where the radius $R$ depends on the angle $\theta$,  can be described  by the surface (\ref{surface})
where 
\begin{eqnarray}
f_{1}(\theta,s)=R(\theta,s)\cos \theta, \qquad f_{2}(\theta,s)=R(\theta,s)\sin\theta.  \qquad 
\end{eqnarray}
Notice that in this case   the cross-section area is given by
\begin{eqnarray}
a(s)=\frac{1}{2}\int_{0}^{2\pi} R^{2}(\theta,s)d\theta. \label{ihc}
\end{eqnarray}

Some geometric quantities as the area of a surface can be written in terms of the first fundamental form, which  is constructed with  the inner product on the tangent space of a surface  as follows    \cite{docarmo}
\begin{eqnarray}
 g =
\left( 
\begin{array}{rrrr}
E &  F\\
F & G \\
\end{array}
\right), \label{fff}
\end{eqnarray}
where 
\begin{eqnarray}
E&=& \frac{ \partial \vec \Sigma(\theta,s)}{\partial s}  \cdot \frac{ \partial \vec \Sigma(\theta,s)}{\partial s},\\
G&=& \frac{ \partial \vec \Sigma(\theta,s)}{\partial \theta} \cdot \frac{ \partial \vec \Sigma(\theta,s)}{\partial \theta},  \\
F&=& \frac{ \partial \vec \Sigma(\theta,s)}{\partial s}\cdot \frac{ \partial \vec \Sigma(s,\theta)}{\partial \theta}. 
\end{eqnarray}

Now, let  us  remember that the curvature $\kappa(s)$ at a point $P$ of the curve $\vec \gamma(s)$ is defined as the inverse of the radius  of the osculating circle at $P,$ see Ref. \cite{spivak}. Then, if the radius  of  the osculating circle is small, the surface (\ref{surface}) describes a cable with a big curvature. In addition, notice that  when the radius  of  the osculating circle is smaller  than the radius of the cable, the cable surface touches itself.  In the literature there are not reported axons with big curvature, then in this paper we suppose that at each point of the curve $\vec \gamma(s)$ the radius of  the osculating circle is  larger than the radius of the cable. Namely,  at each point of the curve $\vec \gamma(s),$ we suppose that the curvature $\kappa(s)$ is smaller than $R^{-1}(s)$ and the following inequality 
\begin{eqnarray}
\kappa(s) R(s) < 1  \label{condition}
\end{eqnarray}
is satisfied. \\

The axon  geometry is important for diverse  physiological processes, such as the voltage propagation. In this respect,
according to Ref. \cite{erick},  when the cable geometry is given by (\ref{surface}) the cable equation is 
\begin{eqnarray}
 \frac{\partial V(s,t)}{\partial t}=\frac{1}{r_{L} c_{M}  \int_{0}^{2\pi} d\theta \sqrt{\det g(\theta,s) }  } \frac{ \partial }{\partial s} \left (a(s) \frac{\partial  V(s,t)  } {\partial s } \right)-   \frac{i_{ion}  }{ c_{M} },\qquad 
\label{cableg} 
\end{eqnarray}
where $a(s)$ is the cable  cross-section and 
\begin{eqnarray}
\sqrt{\det g(\theta,s) }&=& \Bigg[R^{2}(\theta,s) \left(  \frac{\partial R(\theta,s) }{\partial s} -
\tau \frac{\partial R(\theta,s) }{\partial \theta} \right)^{2}\nonumber \\
& & +\left(1-\kappa(s) R(\theta,s) \cos \theta\right)^{2} \left( R^{2}(\theta,s) + \left( \frac{\partial R(\theta,s) }{\partial \theta}\right)^{2}  \right) \Bigg]^{\frac{1}{2}}.\qquad
  \label{saa}
\end{eqnarray} 
Notice that the equation (\ref{cableg}) depends on geometric  quantities as the curvature $\kappa$ and torsion $\tau$ of the cable.\\

Then, in order to study the voltage propagation in a cable with general geometry and anomalous diffusion, 
we can employ  the equations (\ref{frac1}) and (\ref{cableg}) to  propose a generalized fractional cable equation as follows
\begin{eqnarray}
\frac{\partial V(s,t)}{\partial t} = 
\beta_{\nu}  D_{\nu t} \Bigg[\frac{1}{r_{L} c_{M}  \int_{0}^{2\pi} d\theta  \sqrt{\det g(\theta,s) }} \frac{ \partial }{\partial s} \left (a(s) \frac{\partial  V(s,t)  } {\partial s } \right)
 -   \frac{i_{ion}  }{ c_{M} }\Bigg],
\label{fce} 
\end{eqnarray}
where 
 \begin{eqnarray}
D_{\nu t} = \frac{\partial^{1-\nu} }{\partial t^{1-\nu} }, \quad \nu={\rm constant }, \qquad 0\leq \nu \leq 1,
 \end{eqnarray}
is the Riemann-Liouville fractional operator \cite{loke} and  
$\beta_{\nu} $ is a constant with $(time)^{1-\nu}$ dimensions. 
The voltage in  an infinite cable  has to satisfy the Dirichlet boundary condition and  a finite cable has to satisfy the Neumann boundary condition \cite{ermentrout,ermentrout2}, then the solutions
of the equation (\ref{fce}) should  satisfy  these boundary conditions.\\

In the general case, $i_{ion}$ depends on the voltage and the equation (\ref{cableg}) is a non linear differential equation.  
However,  in the  passive cable model we can take 
\begin{eqnarray}
i_{ion}=\frac{V(s,t)}{r_{M}}.
\end{eqnarray}
Therefore,  the fractional cable equation for the passive cable model with the geometry given by (\ref{surface}) is 
\begin{eqnarray}
\frac{\partial V(s,t)}{\partial t} =
 \beta_{\nu}D_{\nu t}\Bigg[\frac{1}{r_{L} c_{M}  \int_{0}^{2\pi} d\theta  \sqrt{\det g(\theta,s) }} \frac{ \partial }{\partial s} \left (a(s) \frac{\partial  V(s,t)  } {\partial s } \right)
 -   \frac{V(s,t)  }{ r_{M}c_{M} }\Bigg].
\label{fce-p} 
\end{eqnarray}

In the next sections we will study some solutions  of this equation.

\section{A qualitative analysis }
\label{se:gc}

For a non trivial geometry, to find solutions of  the equation (\ref{fce-p}) is a difficult task. 
However, let us  provide a qualitative analysis of this equation. In this respect,  we propose the following voltage
 \begin{eqnarray}
V(s,t)={\cal T}(t)X(s). \label{separa}
 \end{eqnarray}
In this case the equation (\ref{fce-p}) implies the following equations
 \begin{eqnarray}
\frac{1}{r_{L} c_{M}  \int_{0}^{2\pi} d\theta  \sqrt{\det g(\theta,s) }}  \frac{ \partial }{\partial s} \left ( a(s) \frac{\partial  X(s)  } {\partial s }\right) -   \frac{X(s)  }{ r_{M}c_{M} } &=& -\lambda X(s), 
 \qquad  \label{spa}\\
 \frac{\partial {\cal T}(t) }{\partial t}&=&-\lambda \beta_{\nu}D_{\nu t}{\cal T}(t) , \qquad \label{tempo}
  \end{eqnarray}
where $\lambda$ is a constant. \\

Moreover, if we take 
 \begin{eqnarray}
X(s)=\frac{\psi(s)}{\sqrt{a(s)}}, \label{spa0}
 \end{eqnarray}
the  spacial equation (\ref{spa}) can be written as 
 \begin{eqnarray}
-\frac{\partial^{2} \psi (s)}{\partial s^{2} }+ U(s) \psi(s)=0,\label{spa1}
\end{eqnarray}
where 
\begin{eqnarray}
U(s)=   \frac{r_{L} c_{M}  \int_{0}^{2\pi} d\theta  \sqrt{\det g(\theta,s) }} {a(s) }\left( \lambda -\frac{1}{r_{M} c_{M}}\right) -\frac{1}{2} \left( \frac{\left( \frac{d a(s)}{ds} \right)^{2}}{2 a^{2}(s) }
 -\frac{ \frac{d^{2} a(s)}{ds^{2}} }{a(s)}\right) . 
 \end{eqnarray}
Observe  that in this last equation the  parameters $\nu$ and $\beta_{\nu}$ do not appear. In fact, this  spacial  equation   
is the same spacial equation which appears  in the  non fractional  case \cite{erick}. As well, observe that the parameter $\lambda$ does not 
 depend on $\nu$  neither on $\beta_{\nu}.$\\

Furthermore, the Laplace transform of the temporal equation (\ref{tempo})  implies 
 \begin{eqnarray}
\tilde {\cal T}_{\nu}(\zeta )= {\cal T}(0) \frac{\zeta ^{\nu -1} }{\left(  \zeta ^{\nu} +  \beta_{\nu} \lambda  \right)  },\label{ltn}
\end{eqnarray}
where $\tilde {\cal T}_{\nu}(\zeta )$ is the Laplace transform of the function ${\cal T}_{\nu}(t).$ The inverse Laplace transform of the function (\ref{ltn})  is given by \cite{loke}
 \begin{eqnarray}
{\cal T}_{\nu }(t)= {\cal T}(0) E_{\nu, 1} \left( -  \beta_{\nu} \lambda  t^{\nu} \right), \label{tem}
\end{eqnarray}
where 
 \begin{eqnarray}
E_{\nu,\Lambda}(z)=\sum_{n\geq 0}\frac{z^{n}}{\Gamma (\nu n +\Lambda  )} \label{fmt}
\end{eqnarray}
 is the Mittag-Leffler type function \cite{loke}. Notice that if  $\nu=1$  we obtain the usual solution 
 \begin{eqnarray}
{\cal T}_{1}(t)= {\cal T}(0)e^{-\lambda   t}.\label{mita}
\end{eqnarray}
In addition, if  $\nu=0$  we obtain 
 \begin{eqnarray}
{\cal T}_{0}(t)= {\cal T}(0) \sum_{n\geq 0} (-\lambda \beta_{0})^{n}, \label{serie}
\end{eqnarray}
which is a constant. We can observe that the equation (\ref{serie}) only makes sense if
\begin{eqnarray}
\beta_{0} \lambda<1, \label{ine}
\end{eqnarray}
in this case we get 
 \begin{eqnarray}
{\cal T}_{0}(t)=\frac{  {\cal T}(0)}{1+\beta_{0} \lambda}. \label{stronger}
\end{eqnarray}
In the section  IV, we can see that, for realistic parameter for cylindrical cable, $\lambda$ is the order of $10^{-3}(sec)^{-1}.$ Then,  the inequality (\ref{ine}) is satisfied if  $\beta_{0}$  is  smaller than $10^{3} sec.$\\

Notice that the strongest fractional effect is obtained  when $\nu$ is close to  zero and  in this case the function ${\cal T}_{\nu}(t)$ is close to the constant (\ref{stronger}), which decreases  when $\beta_{0}$ increases.  Then, in order to model a system where the voltage decreases, we should 
take $\beta_{0}$   bigger than one, but satisfying the inequality (\ref{ine}). If we take $\beta_{0}$ close to zero,   the voltage does not decrease.\\ 

Now, remember  that  the usual case is  obtained with  $\nu=1$ and  $\beta_{1}=1.$ Additionally,  when $\nu$ is close to zero,  $\beta_{\nu}$ should  reach its maximum value, in fact $\beta_{0}$ should be  bigger than one. Thus, for a system where the voltage decreases,    we can suppose that $\beta_{\nu}$ increases when $\nu$ decreases. \\

\begin{figure}[b]
 \begin{center}
        \includegraphics[width=.7\textwidth]{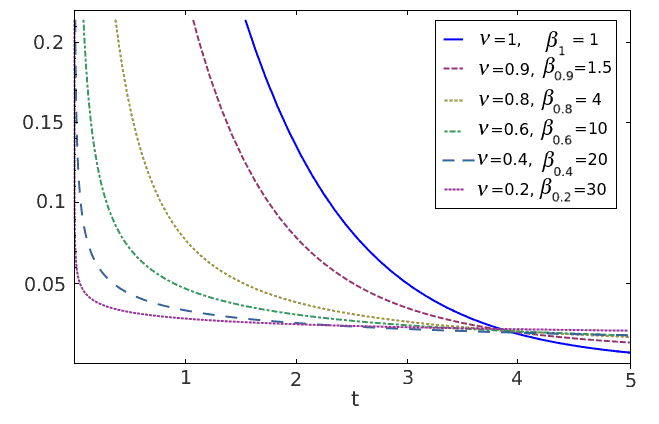}
      \end{center}  
        \caption{   Mittag-Leffler with different values of $\beta_{\nu}$ and $\nu.$  }\label{fig:mittag}
\end{figure}

Fig. \ref{fig:mittag}  shows the function (\ref{tem}) for different $\beta_{\nu}$ and $\nu$ values. 
In this we can see that the function (\ref{tem}) decreases when  $\beta_{\nu}$ increases and  $\nu$ decreases.\\

Then, using the functions (\ref{tem}) and (\ref{spa0}), we found  the following  voltage  
 \begin{eqnarray}
V(s,t)= \frac{{\cal T}(0) }{\sqrt{a(s)} } E_{\nu, 1} \left( -  \beta_{\nu} \lambda  t^{\nu} \right)\psi (s), \label{separa1}
 \end{eqnarray}
 where the function $\psi(s)$ satisfies the equation (\ref{spa1}).  From the  equation (\ref{separa1}) we can see that when the cable  cross-sectional  area $a(s)$  
 increases the voltage  decreases. As well,  when $\beta_{\nu}$ increases and $\nu$ decreases the voltage decreases.  \\

Therefore, this qualitative analysis   suggests   that in an axon with transport or geometrical defects  the voltage decreases. \\

\section{Cylindrical cable with constant radius}
\label{se:ccf}

When the sectional area is  a constant, that is $R(s)=R_{0}=$constant,    the equation (\ref{saa}) does not depend on the curvature neither on  the torsion of the cable. Indeed, in this case the equation (\ref{fce-p}) becomes 
\begin{eqnarray}
\frac{\partial V(s,t)}{\partial t}= \beta_{\nu} D_{\nu t} \left( \frac{R_{0}}{2c_{M} r_{L} }  \frac{\partial^{2}  V(s,t)  } {\partial s^{2} }-   \frac{V(s,t)  }{ r_{M} c_{M} }\right),\qquad 
\label{cablec} 
\end{eqnarray}
which is  equivalent to the fractional cable equation for a straight cylindrical cable (\ref{frac1}). However, observe that the equation (\ref{cablec}) depends on the arc length parameter (\ref{affine}) instead of the lab frame coordinate. This shows that the natural variables for the voltage are given by geometric quantities of the cable.\\

For these geometries, in the finite cable case,  the solution of the equation (\ref{cablec}) is given by
\begin{eqnarray}
 V(s,t)=  \sum_{n\geq 0} b_{n} 
 E_{\nu,1}\left[ -\beta_{\nu} \left(  \frac{n\pi R_{0}}{2c_{M} r_{L} l} +\frac{1}{r_{M} c_{M}}\right) t^{\nu} \right]  \cos n\pi  \frac{ s}{l} ,\quad \label{fvoltage-cyl}
\end{eqnarray}
where $l$ is the length of the cable and $E_{\nu,\Lambda}(z)$
is the Mittag-Leffler  function (\ref{fmt}).\\

\begin{figure*}
        \includegraphics[width=1\textwidth]{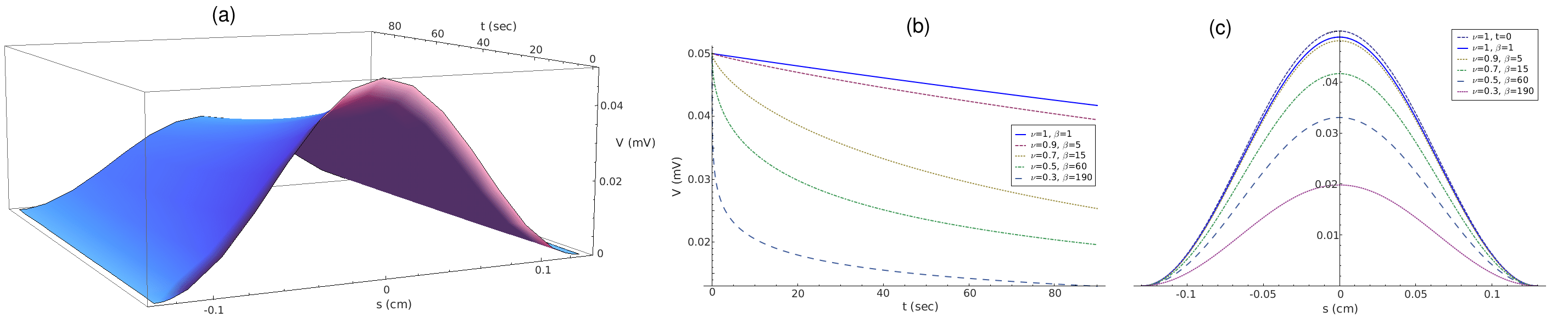}
        \caption{    (a)  Voltage for the cylindrical cable,  with  $\nu=0.7$ and $\beta_{0.7}=15 \;  (sec)^{0.3}.$  (b) Voltage vs $t$ in $s=0$ with  different values of $\beta_{\nu}$ and $\nu.$ (c) Voltage vs $s$ at time $t=12\; sec$ for different values of $\beta_{\nu}$ and $\nu.$ Parameter values used for simulations correspond to realistic dendritic parameters as in \cite{poznanski1}: $c_{M} =1\; mF/cm^2, r_{M} =3000\; \Omega\; cm^2, r_{L} =100 \; \Omega \; cm, R_{0} = 10^{-4} \; cm.$ The initial condition is given by (\ref{inicial-el}). }
\label{fig:cyl-frac-e}
\end{figure*}

Fig.  \ref{fig:cyl-frac-e} shows the fractional cable equation solution for a cylindrical cable for different $\nu$ and $\beta_{\nu}$ values and the 
initial condition 
\begin{eqnarray}
V(s,0)=A\left(1+\cos\left(\frac{s \pi }{l}\right)  \right), \label{inicial-el}
\end{eqnarray}
where $A=0.05 mV\cdot cm ^{\frac{1}{2}}$ and $l=0.13   cm.$  In this figure, we can see that the voltage decreases when  $\beta_{\nu}$ increases and  $\nu$ decreases.  Then,  anomalous diffusion implies 
that voltage decreases.\\

To obtain an exact solution of the equation (\ref{cablec})  is a difficult task for an   arbitrary initial condition.
Then,  in order to study this case  we employ a numerical method. For integer derivatives, spacial and temporal, we use  a second order finite differences method. Moreover, for the temporal fractional derivatives we use a second order scheme taken from the Fractional Integration Toolbox \cite{Ma13}. The mesh size was chosen as follows: first, we begin with 1024 points along the s-axis and 100 points at the time. These numbers were increasing until the difference between two successive solutions was almost null. The number of spatial and temporal points used in the simulations are  shown in  Table \ref{tab:points}. The system is solved using the Gauss-Seidel iterative method, with a tolerance of $10^{-10}$.  In addition,  we impose   the   Neumann boundary  conditions \cite{ermentrout} 
\begin{eqnarray}
\frac{\partial V(s_0,t)}{\partial s} =\frac{\partial V(s_{n_s},t)}{\partial s} = 0. 
\end{eqnarray}

\begin{table}[b]%The best place to locate the table environment is directly after its first reference in text
\caption{\label{tab:points}%
Number of points, $n_s$ and $n_t$, and number of times of the refinement.
}
\begin{center}
\begin{tabular}{| p{3cm} | p{3cm} | p{3cm} |} \hline
Refinement &$n_{s}$ &  $n_{t}$ \\ \hline
First time   & 1024   &  100  \\ \hline
 
   Second time  & 2048   &  500  \\ \hline
  
   Third time   & 4096   &  2000  \\ \hline
\end{tabular}
\end{center}
\end{table}

\begin{table}[b]%The best place to locate the table environment is directly after its first reference in text
\caption{\label{tab:points1}%
Numerical values of $\nu$ and $\beta_{\nu}.$ 
}
\begin{center}
\begin{tabular}{| p{4cm} | p{4cm} | } \hline
\textrm{$\nu$}&
\textrm{$\beta_{\nu}$ }\\ \hline

1  \qquad  &  1  \\ \hline
 
  0.9  \qquad  &  1.5  $(sec)^{0.1}$ \\ \hline
 
  0.7 \qquad   &  4\,    $\;\; (sec)^{0.3}$\\ \hline
   
 0.5 \qquad  &  16 $\; (sec)^{0.5}$  \\ \hline
  
  0.3 \qquad   &  37 $\;  (sec)^{0.7}$ \\  \hline
\end{tabular}
\end{center}
\end{table}

\begin{figure*}
        \includegraphics[width=1\textwidth]{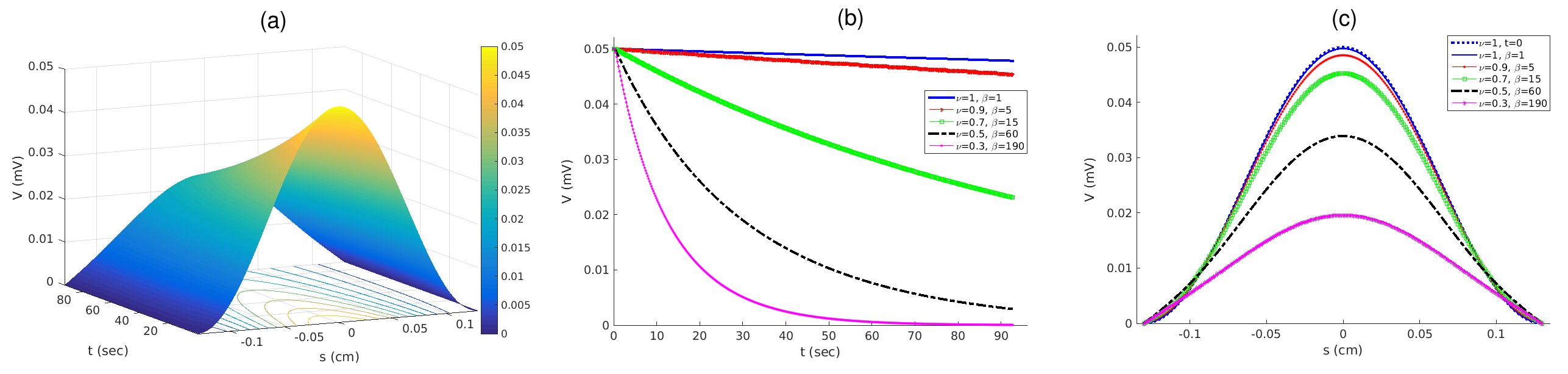}
        \caption{   (a)    Voltage for the cylindrical cable,  with  $\nu=0.7$ and $\beta_{0.7}=15 \;  (sec)^{0.3}.$  (b) Voltage vs $t$ in $s=0$ with  different values of $\beta_{\nu}$ and $\nu.$ (c) Voltage vs $s$ at time $t=12\; sec$ for different values of $\beta_{\nu}$ and $\nu.$ Parameter values used for simulations correspond to realistic dendritic parameters as in \cite{poznanski1}: $c_{M} =1\; mF/cm^2, r_{M} =3000\; \Omega\; cm^2, r_{L} =100 \; \Omega \; cm, R_{0} = 10^{-4} \; cm.$ The initial condition is given by (\ref{inicial-el}). }
\label{fig:cyl-frac-n}
\end{figure*}

Fig.  \ref{fig:cyl-frac-n}  shows the numerical solution of  the equation (\ref{cablec})  for a cylindrical cable for different $\nu$ and $\beta_{\nu}$ values and the initial condition (\ref{inicial-el}). This figure shows that the numerical solution is close to the analytical solution.\\ 

\begin{figure*}
        \includegraphics[width=1 \textwidth]{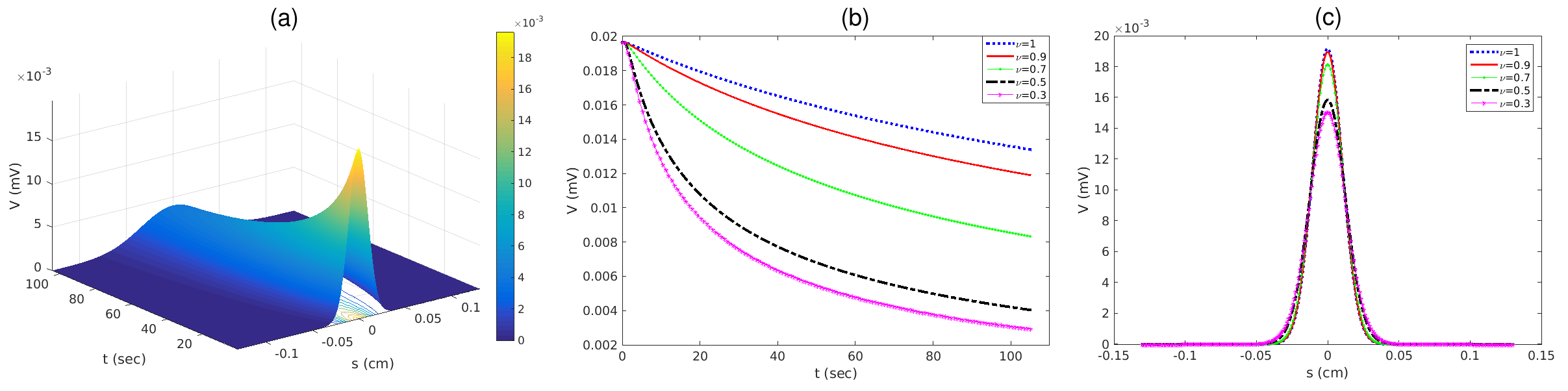}
        \caption{    (a)  Voltage for the cylindrical cable,  with  $\nu=0.5$ and $\beta_{0.5}=16 \;  (sec)^{0.5}.$  (b) Voltage vs $t$ in $s=0$ with  different values of $\beta_{\nu}$ and $\nu.$ (c) Voltage vs $s$ at time $t=7\; sec$ for different values of $\beta_{\nu}$ and $\nu.$ Parameter values used for simulations correspond to realistic dendritic parameters as in \cite{poznanski1}: $c_{M} =1\; mF/cm^2, r_{M} =3000\; \Omega\; cm^2, r_{L} =100 \; \Omega \; cm, R_{0} = 10^{-4} \; cm.$ The initial condition is given by (\ref{inicial}). }
\label{fig:cyl-frac}
\end{figure*}

A more realistic initial condition is given by the function  
\begin{eqnarray}
V(s,0)=\frac{A}{\sqrt{2\pi\sigma}}e^{-\frac{s^2}{2\sigma^2}}, \label{inicial}
\end{eqnarray}
 where $A=0.00128 mV\cdot cm ^{\frac{1}{2}}$ and $\sigma=0.004   cm.$
 Fig.  \ref{fig:cyl-frac} shows the numerical  solution of equation (\ref{cablec})  for different $\nu$ and $\beta_{\nu}$ values and the initial condition 
 (\ref{inicial}). The numerical values of $\nu$ and $\beta_{\nu}$  used in this simulation  are  shown in  Table \ref{tab:points1}. \\

Notice that  both  the exact and the numerical solutions of the equation (\ref{cablec}) provide   a voltage which decreases when  $\beta_{\nu}$ increases and  $\nu$ decreases.  Then,  anomalous diffusion implies 
that voltage decreases. It is worth mentioning that some experimental studies  show  that  in brain tumours there is    anomalous diffusion 
\cite{brain-t1,brain-t2,brain-t3}. As well,  in different neurodegenerative diseases disruption axonal transport are reported \cite{transport1,transport2,transport3}. In these cases, the fractional 
cable equation implies that the voltage decreases. \\

 In the next section we will study the equation (\ref{fce-p})   with non constant radius. \\

\section{Cables with swellings  }
\label{se:ns}

It can be shown that  when a cable has  vanished curvature  and  the cable radius is  given by
\begin{eqnarray}
R(s) =  R_{0}\left( 1 + \alpha_{1}\sin\alpha_{2} s \right) \label{sin}
 \end{eqnarray}
or 
\begin{eqnarray}
R(s) =  R_{0}\left( 1 + \alpha_{1}\sin^{2}\alpha_{2} s \right) \label{sin2},
 \end{eqnarray}
the spatial equation (\ref{spa1}) is similar to the spatial  equation for the straight cylindrical cable. For this reason,  the voltage in a cable with 
radius (\ref{sin}) or (\ref{sin2}) is similar to the  voltage in the straight cylindrical cable \cite{erick}. Hence, in the fractional case, when a  cable has  vanished curvature  and  the radius is given by (\ref{sin}) or (\ref{sin2}) the voltage will be similar to voltage  (\ref{fvoltage-cyl}). \\

\subsection{Circular cross-section}
\label{se:cics}

The cable equation (\ref{fce-p})  is hard to solve for a  general cable geometry. However, for  some cases  this equation can be simplified.
 For instance, when the cable has   a circular cross-section, 
namely when $R(\theta,s)=R(s),$  the equation  (\ref{fce-p}) does not depend on 
the torsion of the cable $\tau$ and becomes
{\small \begin{eqnarray}
\frac{\partial V(s,t)}{\partial t}= \beta_{\nu}D_{\nu t}
\left(\frac{\pi \frac{ \partial }{\partial s} \left ( R^{2}(s)  \frac{\partial  V(s,t)  } {\partial s } \right) }{r_{L} c_{M} R(s) \int_{0}^{2\pi} d\theta \sqrt{ ( 1-\kappa(s) R(s) \cos \theta  )^{2} +\left( \frac{dR(s)}{ds}\right)^{2}}  } -   \frac{V(s,t)  }{ r_{M} c_{M} }\right).\qquad 
\label{cableg1} 
\end{eqnarray}
}
In the following subsections we will study cables which model   axons with swellings.

\subsection{Cable with  Gaussian swelling }

\begin{figure*}
        \includegraphics[width=1 \textwidth]{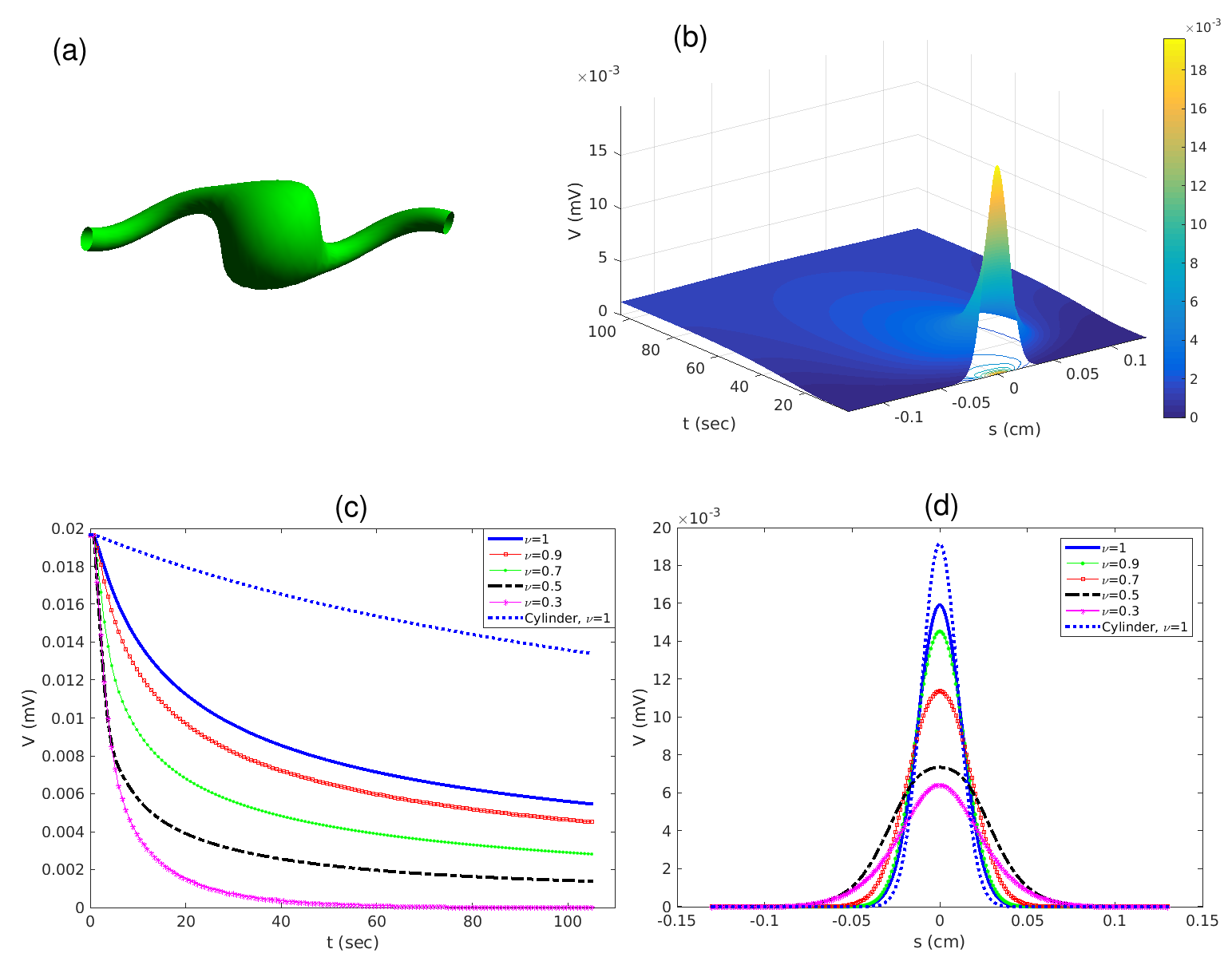}
        \caption{  (a) Cable with geometry (\ref{radius1}). (b)  Voltage for the cable with the radius (\ref{radius1}) and  
         $\nu=0.5, \beta_{0.5}=16\; (sec)^{0.5}.$  (c) Voltage vs $t$ in $s=0$ with  different values of $\beta_{\nu}$ and $\nu.$ (d) Voltage vs $s$ at time $t= 7 \; sec$ for different values of $\beta_{\nu}$ and $\nu.$ Parameter values used for simulations correspond to realistic dendritic parameters as in \cite{poznanski1}: $c_{M} =1\; mF/cm^2, r_{M} =3000\;  \Omega \; cm^2, r_{L} =100\; \Omega  \; cm, R_{0} = 10^{-4}\;  cm, \alpha_{1}=10,\;  
         \alpha_{2}=0.11 \;cm^{-2},\; \alpha_{3}=0\; cm.$ The initial condition is given by (\ref{inicial}).}\label{fig:gaussian}
\end{figure*}

In different neurodegenerative diseases, focal axonal swellings are found such as the Fig.  \ref{fig:gaussian}(a), see for example  \cite{Tr98}.
This  geometry can be modelled with a cable with radius 
\begin{eqnarray}
R(s) =  R_{0}\left( 1 + \alpha_{1} e^{-\alpha_{2}(s - \alpha_{3})^{2}}  \right).  \label{radius1}
 \end{eqnarray}
For this cable geometry, if the initial condition is given by  (\ref{inicial}), the numerical solution of the equation (\ref{cableg1})  
 can be seen in the Fig.  \ref{fig:gaussian}(b), (c) and (d).  In these figures  we can observe that 
 the voltage decreases faster than the voltage of  the cylindrical cable.  In addition, notice that for this cable geometry   
  when $\beta_{\nu}$ increases and $\nu$ decreases   the voltage of 
 the cable decreases faster than when $\beta_{1}=1$ and $\nu=1.$ Then, in an axon with a swelling and anomalous diffusion 
 the voltage strongly decreases.

\subsection{Cable with Gaussian swellings }

In this section we study the numerical solution for the equation (\ref{cableg1}) for  a cable with the radius 
%
%\small{
\begin{eqnarray}
R(s) &=&  R_{0}\left(1 + \alpha_{1} e^{-\alpha_{2}s^{2}}  + \alpha_{1} e^{-\alpha_{2} (s - \alpha_{3})^{2}}
+\alpha_{1} e^{-\alpha_{2}(s - 2\alpha_{3})^{2}}  + \alpha_{1} e^{-\alpha_{2} (s - 3\alpha_{3})^{2}} \right)  \qquad \label{radius2}
 \end{eqnarray}
% }
 % 
 and  with the initial condition  (\ref{inicial}). \\
  
\begin{figure*}
        \includegraphics[width=1\textwidth]{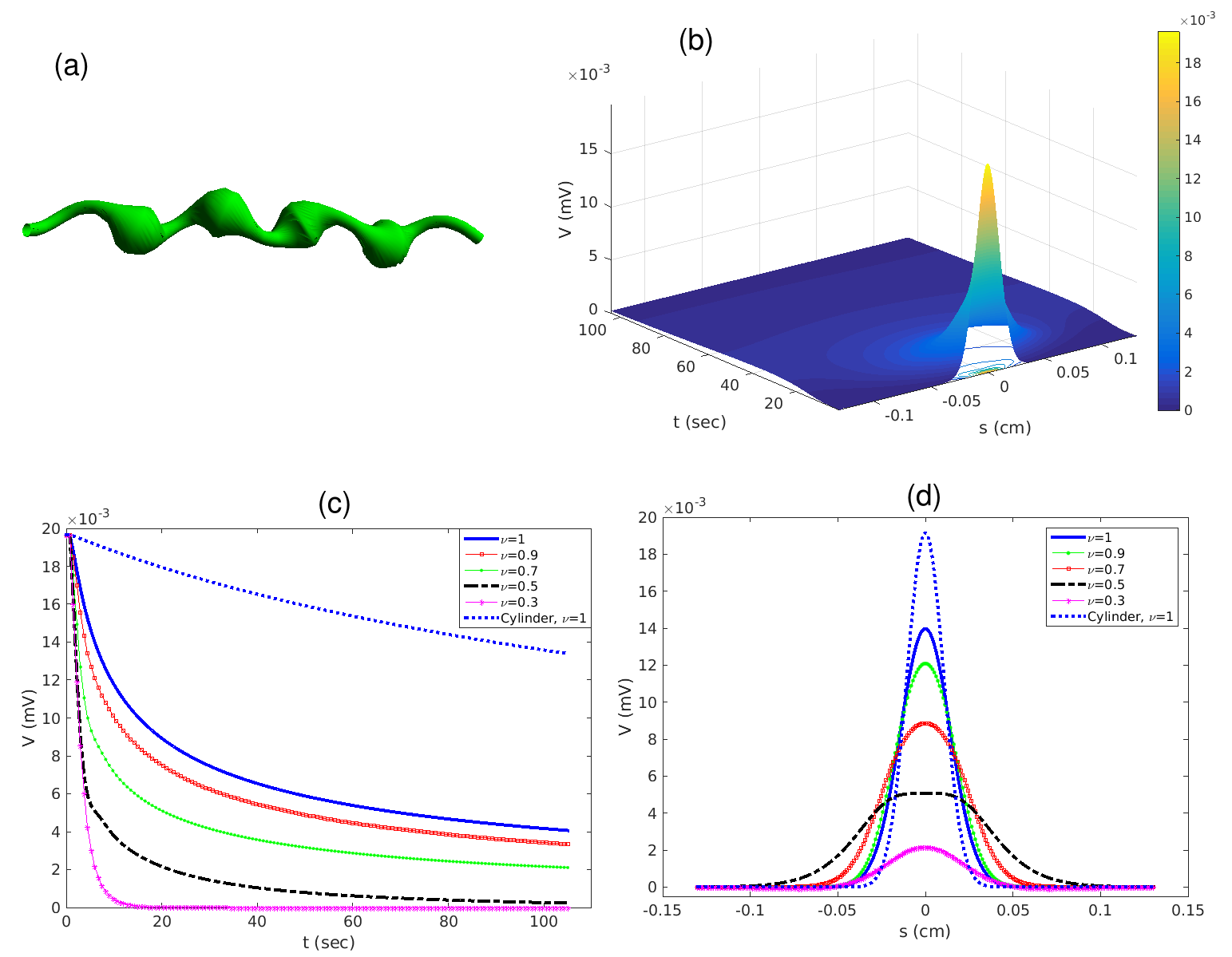}
        \caption{  (a) Cable with geometry (\ref{radius2}). (b)  Voltage for the cable with a gaussian train swellings, radius (\ref{radius2})  and   $\nu=0.5, \beta_{0.5}=16 \, (sec)^{0.5}.$  (c) Voltage vs $t$ in $s=0$ with  different values of $\beta_{\nu}$ and $\nu.$ (d) Voltage vs $s$ at time $t=7 \, sec$ for different values of $\beta_{\nu}$ and $\nu.$ Parameter values used for simulations correspond to realistic dendritic parameters as in \cite{poznanski1}: $c_{M} =1mF/cm^2, r_{M} =3000\, \Omega \, cm^2, r_{L} =100\, \Omega \, cm, R_{0} = 10^{-4}\, cm,\; \alpha_{1}=10,\;  
         \alpha_{2}=0.11 \;cm^{-2},\; \alpha_{3}=0.06 \,cm.$ The initial condition is given by (\ref{inicial}). }\label{fig:gaussiant}
\end{figure*}

 A cable with this geometry can be seen in the figure \ref{fig:gaussiant}(a). It is worth mentioning that axons with this geometry have been reported in different studies \cite{Sh98}. The numerical solution of  the equation (\ref{cableg1}) for this  cable radius  can be observed  in Fig.  \ref{fig:gaussiant}(b), (c) and (d).  In this case  the voltage decreases faster than the voltage of  the cylindrical cable.  Moreover,   when $\beta_{\nu}$ increases and $\nu$ decreases,   the voltage  decreases faster than non fractional voltage  ($\beta_{1}=\nu=1$).  In addition, we can see that the voltage in a cable with this geometry 
 decreases faster than the voltage in a cable with the geometry (\ref{radius1}).

\subsection{Amorphous swelling }
\label{se:ceas}

\begin{figure*}
        \includegraphics[width=1\textwidth]{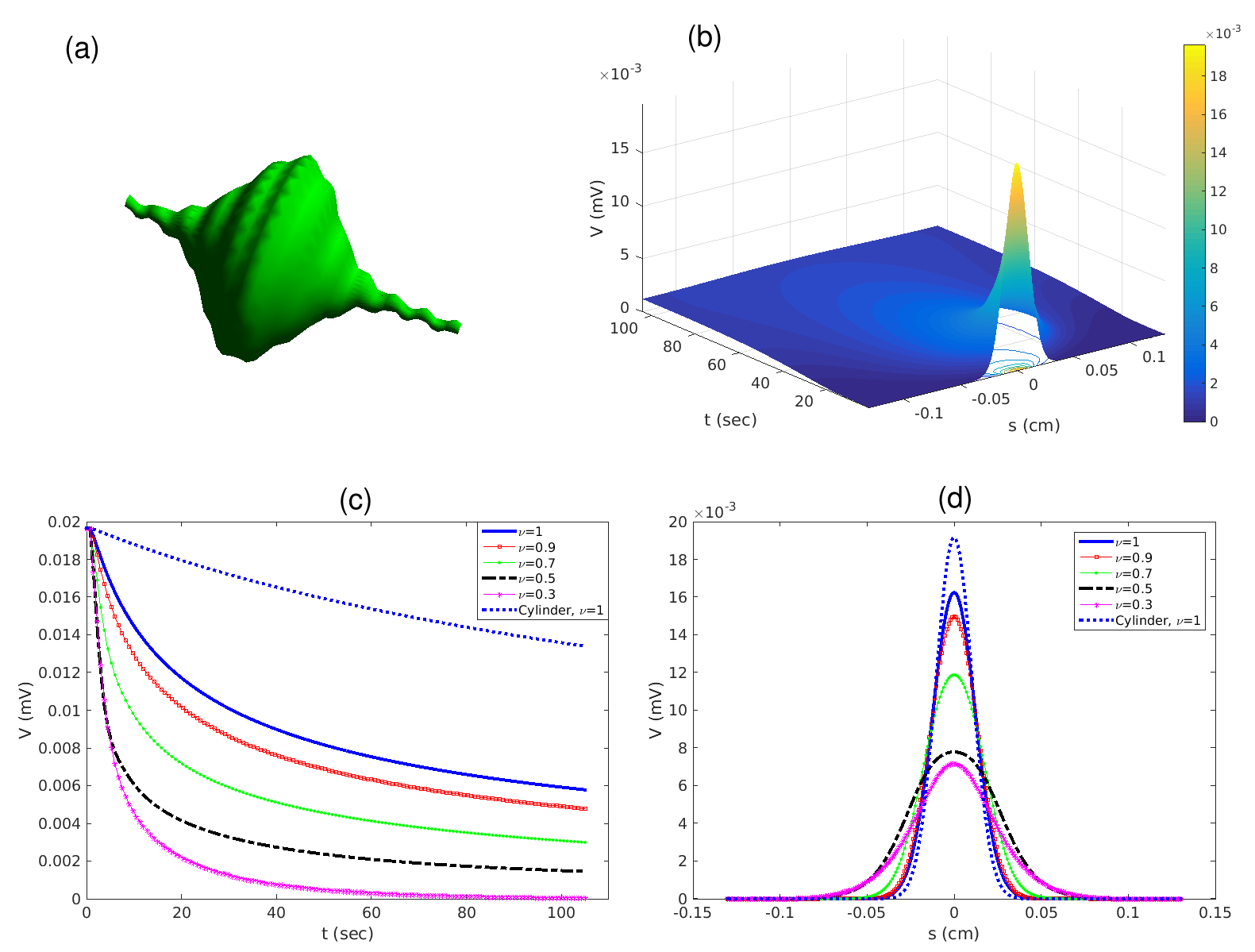}
 \caption{ (a) Cable with geometry (\ref{amorphusc2}). (b)  Voltage for the cable with the radius (\ref{amorphusc2}) and   $\nu=0.5, \beta_{0.5}=16 \; (sec)^{0.5}.$   (c) Voltage vs $t$ in $s=0$ with  different values of $\beta_{\nu}$ and $\nu.$ (d) Voltage vs $s$ at time $t=7\; sec$ for different values of $\beta_{\nu}$ and $\nu.$ Parameter values used for simulations correspond to realistic dendritic parameters as in \cite{poznanski1}: 
 $c_{M} =1\;  mF/cm^2, r_{M} =3000\; \Omega \; cm^2, r_{L} =100 \; \Omega \; cm, R_{0} = 10^{-4}\; cm,\;  \alpha_{1}=10,\;  
         \alpha_{2}=0.11 \;cm^{-2},\; \alpha_{3}=0 \,cm, \alpha_{4}=10, \; \alpha_{5}= 0.11\; cm^{-1}.$ The initial condition is given by (\ref{inicial}).   }\label{fig:gaussianamo1}   
 \end{figure*}

A more realistic model for an axon with swelling is given by Fig.  \ref{fig:gaussianamo1}(a). 
This  geometry can be described by a cable with an  amorphous swelling with radius 
\begin{eqnarray}
R(\theta,s) = R_{0}\left(1 +  \alpha_{1}e^{-\alpha_{2}(s - \alpha_{3}) ^{2} } + \alpha_{4} \sin\theta  \cos \alpha_{5}s \right). \label{amorphusc2}\qquad 
\end{eqnarray}
The numerical solution of  the voltage  for this cable can be seen in Fig. \ref{fig:gaussianamo1}(b), (c) and (d).  This  voltage
is different from  the voltage for the cable with radius (\ref{radius1}). Then, geometric inhomogeneities affects the voltage propagation in a cable.

\section{Summary}
\label{se:s}
In different neurodegenerative diseases  such as multiple sclerosis, Alzheimer's disease and  Parkinson's disease are reported  abnormal accumulations of proteins and organelles which generate  disruption axonal transport. In addition, recently different experimental studies have found  anomalous diffusion in brain tissues. Notably this diffusion is expressed through fractional calculus. Another hallmark feature  of   some neurodegenerative diseases  is given by axonal discrete  swellings along the axons. \\

In order to study the voltage propagation in a cable with  both of these hallmark features of the  neurodegenerative diseases,
we proposed a  fractional cable equation with non trivial  geometry. This equation depends  on geometric quantities such 
as the  curvature and torsion of the cable,  as well as the fractional parameters  
$\beta_{\nu}$ and  $\nu.$  It is worth  mentioning that the parameter  $\beta_{\nu}$ depends on $\nu.$  In this respect, we  showed that  with regard to model a system where the voltage decreases, we should  suppose that $\beta_{\nu}$ increases when $\nu$ decreases. Furthermore, in this new cable equation the strongest fractional effect is obtained when $\nu$ is close to zero.\\
 
 For  a straight cylinder  with a constant radius we showed that  the voltage decreases  when the fractional effect  increases. Notice that in this case the cable does not have swellings. Then, if there is an  abnormal accumulation of proteins and organelles in an axon,  the diffusion can be hindered and become to anomalous diffusion. In this case  
 our results suggest that  the voltage decreases in axons. Indeed, when the fractional effect is strong, the voltage may  be blocked.\\

In addition, cables with swellings and anomalous diffusion were studied. Regarding this,  we studied cable  geometries similar to some axons reported in the literature. For all these cable geometries,  we found that   when the fractional effect increases,  the voltage  decreases.
Furthermore, we found that the voltage  dramatically decreases when the cable has a big swelling  or has many swellings. Then, these results also suggested that in axons with swellings and  abnormal accumulations of proteins and organelles the voltage decreases and may be blocked.\\

How axonal transport defects and deformed geometry  of  axons   are related each other is an important problem. In order to study this problem, in our model the geometry quantities of the cable should be related with the parameters  $\nu$ and $\beta_{\nu}.$ 
In a future work  the above problem and  the active case, where non linear interactions play an important role, will be studied. 

\begin{center}
{\bf Acknowledgments}
\end{center}

This work was supported in part by  CONACyT-SEP 47510318 (J.M.R.).  We are grateful to referees for providing valuable
comments.

\end{document}